\documentclass[conference]{IEEEtran}
\ifCLASSINFOpdf
\else
\fi
\usepackage{graphicx}

\begin{document}
%
\title{A PSO Approach for Optimum Design of Multivariable PID Controller for nonlinear systems}

\author{\IEEEauthorblockN{Taeib Adel}
\IEEEauthorblockA{Research Unit on Control,\\ Monitoring and Safety of Systems (C3S)\\High School ESSTT\\
Email: taeibadel@live.fr}
\and
\IEEEauthorblockN{Ltaeif Ali}
\IEEEauthorblockA{Research Unit on Control,\\ Monitoring and Safety of Systems (C3S)\\High School ESSTT\\
Email: ltaief24@yahoo.fr}
\and
\IEEEauthorblockN{Chaari Abdelkader}
\IEEEauthorblockA{Research Unit on Control,\\ Monitoring and Safety of Systems (C3S)\\High School ESSTT\\
Email: nabile.chaari@yahoo.fr}}


%


\maketitle

\begin{abstract}
The aim of this research is to design a PID Controller using particle swarm optimization (PSO) algorithm for multiple-input multiple output (MIMO)  Takagi-Sugeno fuzzy model. The conventional gain tuning of PID controller (such as Ziegler-Nichols (ZN) method) usually produces a big overshoot, and therefore modern heuristics approach such as PSO are employed to enhance the capability of traditional techniques. However, due to the computational efficiency, only PSO will be used in this paper. The results show the advantage of the PID tuning using PSO-based optimization approach.
\end{abstract}


%
\IEEEpeerreviewmaketitle

\section{Introduction}
PID control, which is usually known as a classical output feedback control for SISO systems,
has been widely used in the industrial world [1] and [2]. The
tuning methods of PID control are adjusting the proportional, the integral and the derivative
gains to make an output of a controlled system track a target value properly. several researchers focus on multiple-input–multiple output MIMO control systems. Because many industrial
processes are MIMO systems which need MIMO control techniques to improve performance, though they
are naturally more difficult than SISO systems. As we know, MIMO PID controller design has developed over
a number of years. Luyben (1986) proposed a simple tuning method for decentralized PID controllers in
MIMO system from single-loop relay tests [3]. Yusof and Omatu (1993) presented a multivariable self-tuning PID
controller based on estimation strategies [4]. Wang et al. (1997) proposed a tuning method for fully
cross-coupled multivariable PID controller from decentralized relay feedback test to find the critical oscillation frequency of the system by first designing the diagonal elements of multivariable PID controller independent of off-diagonal ones [5]. Recently, the computational intelligence has proposed particle swarm optimization (PSO) [6,7] as opened paths to a new generation of advanced process control. The PSO algorithm, proposed by Kennedy and Eberhart [6] in 1995, was an evolution computation technology
based on population intelligent methods. In comparison
with genetic algorithm, PSO is simple,easy to realize and has
very deep intelligent background. It is not only suitable for scientific
research, but also suitable for engineering applications in
particular. Thus, PSO received widely attentions from evolution
computation field and other fields. Now the PSO has become a
hotspot of research. Various objective functions based on error
performance criterion are used to evaluate the performance of
PSO algorithms.

In this paper, a scheduling PID tuning parameters using particle swarm optimization strategy for  MIMO nonlinear systems . This paper has been organized as follows: In section 2, a brief
review of the TS fuzzy model formulation is given. Estimation method of recursive weighted least-squares (RWLS)in section 3. In section 4, PID control systems of multivariable processes. Finally, some conclusions are made in section 5.

%

\section{Takagi-Sugeno fuzzy model of a MIMO process}
Generally, modeling process consists to obtain a parametric model with the same dynamic behavior of the real process. In this section, we are interested to the problem of the MIMO process identification[8]. We consider a MIMO system with $n_{i}$ inputs and $n_{0}$ outputs.
The MISO models are a input-output NARX (Non linear Auto Regressive
with eXogenous input) defined by:
\begin{equation}\label{3}
  y_{l}(k+1)=f_{l}(x_{l}(k))\,\,\,\,\,\,\,\,\,\,l=1,2,...,n_{0}
\end{equation}
With the regression vector represented by:
\begin{equation}\label{2}
\begin{array}{c}
  x_{l}(k)=[\{y_{1}(k)\}_{0}^{n_{yl1}},\{y_{2}(k)\}_{0}^{n_{yl2}},...,\{y_{n0}(k)\}_{0}^{ln_{yln_{0}}},\\
  \,\,\,\,\,\,\,\,\,\,\,\,\, \{u_{1}(k)\}_{n_{dl1}}^{n_{ul1}},\{u_{2}(k)\}_{n_{dl2}}^{n_{ul2}},...,\{u_{n_{i}}(k)\}_{n_{dln_{i}}}^{n_{ud1n_{i}}}]
\end{array}
\end{equation}
$n_{y}$ and $n_{u}$ define the number of delayed outputs and inputs respectively. $n_{d}$
is the number of pure delays. $n_{y}$ is a $n_{0}*n_{0}$ matrix and $n_{u}$, $nd$ are $n_{0}*n_{i}$
matrices. $f_{l}$ are unknown non linear functions.
MISO models are estimated independly [9], so, to simplify the notation, the
output index $l$ is omitted and we will be interested only in the multi-input,
mono-output case.
The Takagi-Sugeno MISO rules are estimated from the system input- output data[15]. The base rule contains $r$ rules of the following form:
\begin{equation}\label{eq3}
\begin{array}{l}
\textbf{R}_{i}: If y_{1}(k)\,\,\,is \,\,\,\textbf{A}_{i1} \,\,\,and\,\,\, if\,\,\, u_{nu}(k-n+1)\,\,\,is\,\,\, \textbf{A}_{_{ino}}\,\,\,\\\textbf{then}\,\,\,
y_{i}(k+1) = \sum\limits _{j=1}^{n}a_{ir} y_{i}(k-j+1) + \sum\limits  _{j=1}^{n}b_{ij} u_{i}(k-j+1)\\
+\sum\limits _{l=1}^{nu}\sum\limits _{j=1}^{n}b_{ilj}u_{l}(k+j-1)+c_{i}\,\,\,\,\,$i=1,2..,r$
\end{array}
\end{equation}

 \section{Estimation method of recursive weighted least-squares (RWLS)}
For nonlinear systems the online adaptation is necessary to obtain a model able to continue the system in its evolution. The system described by relation (4) can also be rewritten as:
\begin{equation}\label{eq5}
y_{i}(k) = \theta ^t_{i} \varphi_{i}(k-1)
\end{equation}
 with $\theta$ being a system parameter vector and $\varphi$ a regression vector. It should be noted that the system (5) is in general nonlinear but it is linear with respect to its unknown parameter vectors. Based on parameterizations (\ref{eq5}), the identification algorithm giving estimates $\widehat{\theta}(k)$ of $\theta(k)$ can be obtained using the RWLS.\\
  We define:
\begin{equation}\label{eq6}
\begin{array}{c}
 \varphi_{i} ( {k - 1} ) =[\mu_{i}\,y_{i}(k - 1)...\,\mu_{ij} y_{i}(k -n)\\
 \;\;\;\;\;\;\;\;\;\;\;\;\;\;\;\;\;\;\;\;\;\;\;\;\;\mu_{i}u_{i}(k - 1)...\,\mu_{i} u_{i}(k - n)\,\,\, \mu_{i} ]
\end{array}
\end{equation}
\begin{equation}\label{eq7}
\theta _i  = \left[ {a_{i1} ...a_{in} \,b_{i1} ...b_{in} c_i } \right]
\end{equation}
\begin{equation}\label{eq8}
\varphi_{i} \left( {k - 1} \right) = \left[ {\varphi^t_{i1} \left( {k - 1} \right)\,\varphi^t_{i2} \left( {k - 1} \right)...\varphi _{ir}^t \left( {k - 1} \right)} \right]^t
\end{equation}
\begin{equation}\label{eq9}
 \theta_{i}(k)=\theta_{i}(k-1)+L_{i}(k)[y_{i}(k)-\varphi^{t}(k)\theta_{i}^{t}(k-1)]
\end{equation}
\begin{equation}\label{eq10}
    L_{i}(k)=\frac{P(k-1)*\varphi^{t}(k)}{1/\mu_{ik}+\varphi(k)P(k-1)\varphi^{t}}
\end{equation}
\begin{equation}\label{eq9}
P(k) = P_{i}({k - 1} -L_{i}(k)\varphi(k) P_{i}(k-1)
\end{equation}
for $k=1,...,N, P(k-1)$ is a covariance matrix and L(k) referred to the estimator gain vector. A common choice of initial value is to take $\theta _i(0)=0$ and $P_{i}(0)=\alpha I$ where $\alpha $ is a large number.
\section{PID control systems of multivariable processes}
Consider a multivariable PID control structure as shown in Fig. 1,
where:\\
Desired output vector : $\textbf Y_{d}=[y_{d1},y_{d2},...,y_{dn}]^{T}$.\\ Actual output vector: $\textbf Y=[y_{1},y_{2},...,y_{n}]^{T}$.\\
 Error vector : $\textbf{E}=\textbf {Y}_{d}-\textbf{Y}=[y_{d1}-y_{1},y_{dn}-y_{2},...,y_{dn}-y_{n}]=[e_{1},e_{2},...,e_{n}]^{T}$.\\
 Control input vector : $\textbf{U}=[u_{1},u_{2},...,u_{n}]^{T}$.\\
 $n*n$ multivariable processes:
 \begin{equation}\label{1}
    \textbf{H}(z)=\left| \begin{array}{ccc} h_{11}(z)&...& h_{1n}(z) \\ ...&...&...\\h_{n1}(z)&... &h_{nn}(z) \end{array} \right|
 \end{equation}
$n*n $multivariable PID controller:
\begin{equation}\label{2}
    \textbf{C}(z)=\left| \begin{array}{ccc} c_{11}(z)&...& c_{1n}(z) \\ ...&...&...\\c_{n1}(z)&... &c_{nn}(z) \end{array} \right|
 \end{equation}
 The form of $kij(z)$, for $i, j\in\underline{n}$ and $\underline{n} = \{1, 2, . . . ,n\}$, is given by:
 \begin{equation}\label{8}
    \textbf{C}(z)=kp_{ij}(1+\frac{z}{Ti_{ij}(z-1)}+\frac{(z-1)}{Td_{ij}z})
 \end{equation}
where $kp$ is the proportional gain, Ti is the integral time
constant, and $Td$ is the derivative time constant. It can be
also rewritten (9) as:
\begin{equation}\label{10}
    \textbf{C}(z)=kp_{ij}+\frac{ki_{ij}z}{(z-1)}+\frac{kd_{ij}(z-1)}{z})
 \end{equation}
 where $ki = kp/Ti$ is the integral gain and $kd = kp*Td$ is the
derivative gain. For convenience, let $K=[kp_{ij} ; ki_{ij}$ ; $kd_{ij}]^{T}$
represent the gains vector of $i^{th}$ row and $j^{th}$ column[12].
In the design of a PID controller, the performance criterion
or objective function is first defined based on our desired
specifications and constraints under input testing signal.
Typical output specifications in the time domain are peak
overshooting, rise time, settling time, and steady-state error,
to name a few.
\begin{figure}[thpb]
     \centering
    \includegraphics[scale=0.20]{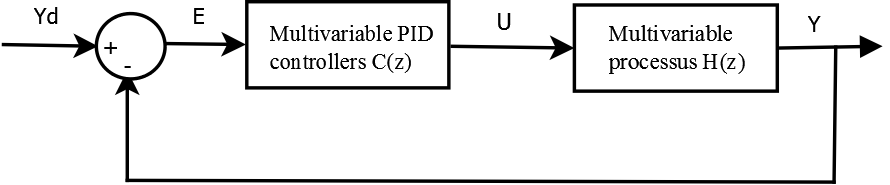}
       \caption{A multivariable PID control system.}
      \label{figure1}
   \end{figure}
Three kinds of performance criteria usually considered in the control design are integral of the Absolute Error (\textbf IAE), integral of Square Error (\textbf ISE) and integral of Time weighted Square Error (\textbf ITSE) which are  given as:
\begin{equation}\label{11}
\textbf{IAE}=|{e_{1}(k)+e_{2}(k)}+...+e_{n}(k)|
\end{equation}
\begin{equation}\label{12}
\textbf{ISE}=\sum {e_{1}(k)} ^2+\sum {e_{2}(k)} ^2+...+\sum {e_{n}(k)} ^2
\end{equation}
\begin{equation}\label{13}
\textbf{ITSE}=\sum k*{e_{1}(k)} ^2+\sum k*{e_{2}(k)} ^2+...+\sum k*{e_{n}(k)} ^2
\end{equation}
Therefore, for the PSO-based PID tuning, these
performance indexes (Eqs. (17)-(19)) will be used as the objective
function. In other word, the objective in the PSO-based
optimization is to seek a set of PID parameters such that the
feedback control system has minimum performance index.
\subsection{Tuning of PID uzing Z-N method}
The first method of Z-N tuning is based on the open-loop
step response of the system. The open-loop system's Shaped
response is characterized by the parameters, namely
the process time constant $T$ and $L$. These parameters are
used to determine the controller's tuning parameters (see TABLE.1).
\begin{table}[h]
\caption{Ziegler-Nichols open-loop tuning parameter}
\label{table_example}
\begin{center}
\begin{tabular}{|c||c||c||c|}
\hline
$\textbf{Controller}$ & kp & Ti=kp/ki & Td=kd/kp \\
\hline
\textbf{P} &T/L & -&0 \\
\hline
\textbf{PI} &0.9(T/L) &L/0.3&0\\
\hline
\textbf{PID}&1.2(T/L) & 2L&0.5L\\
\hline
\end{tabular}
\end{center}
\end{table}
The second method of Z-N tuning is closed-loop tuning
method that requires the determination of the ultimate gain
and ultimate period. The method can be interpreted as a
technique of positioning one point on the Nyquist curve [13].
This can be achieved by adjusting the controller gain (Ku)
till the system undergoes sustained oscillations (at the
ultimate gain or critical gain), whilst maintaining the integral time constant ( Ti ) at infinity and the derivative time
constant (Td) at zero (see TABLE.2).
\begin{table}[h]
\caption{Ziegler-Nichols closed-loop tuning parameter}
\label{table_example}
\begin{center}
\begin{tabular}{|c||c||c||c|}
\hline
$\textbf{Controller}$ & kp & Ti=kp/ki & Td=kd/kp \\
\hline
\textbf{P} &0.5ku& -&0 \\
\hline
\textbf{PI} &0.45ki &1.5kp/Pu&0\\
\hline
\textbf{PID}&0.6ku &2kp/Pu&kpPu/8\\
\hline
\end{tabular}
\end{center}
\end{table}
\subsection{Implementation of PSO-Based PID Tuning}
\subsubsection{Particle swarm optimization (PSO)}
Particle swarm optimization was introduced by Kennedy
and Eberhart by simulating social behavior of birds flocks in (199[10]. The PSO algorithm has been successfully applied to solve various optimization problems
[14]. The PSO works by having a group of m particles. Each particle
can be considered as a candidate solution to an optimization problem
and it can be represented by a point or a position vector
$\textbf{X}_{ij} = [X_{i1}, . . . , X_{id}]$ in a d dimensional search space which keeps on
moving toward new points in the search space with the addition
of a velocity vector $\textbf{V}_{ij} = [V_{i1}, . . . , V_{id}]$ to further facilitate the search
procedure. The initial positions and velocities of particles are random
from a normal population in the interval [0, 1]. All particles move in the
search space to optimize an objective function $\textbf{f}$. Each member
of the group gets a score after evaluation on objective function $\textbf{f}$.
The score is regarded as a fitness value. The member with the highest
score is called global best. Each particle memorizes its previous best
positions. During the search process all particles move toward the
areas of potential solutions by utilizing the cognitive and social
learning components. The process is repeated until any prescribed
stopping criterion is reached. After any iteration, all particles
update their positions and velocities to achieve better fitness values
according to the following:
\begin{eqnarray}
\textbf{V}_{pd}^{t+1}&=&\omega \textbf{V}_{pd}^{t}+c_{1}r_{1}(\textbf{pbest}^{t}-\textbf{X}_{pd}^{t})\\
&+&c_{2}r_{2}(\textbf{gbest}^{t}-\textbf{X}_{id}^{k}), \nonumber \\
\textbf{X}_{pd}^{t+1}&=&\textbf{X}_{pd}^{t}+\textbf{V}_{pd}^{t+1},
\end{eqnarray}
where:\\
        $ t$ is the current iteration number, $\textbf{pbest}_{i}$ is ${pbest}$ of particle $i$, $\textbf{gbest}_{g}$ is ${gbest}$ of the group, $r_{1},r_{2}$ are two random numbers in the interval [0, 1],  $c_{1},c_{2}$ are positive constants and $w$ is the inertia weight,is a parameter used to control the impact of the previous velocities on the current velocity. It influences the tradeoff between the global and local exploitation abilities of the particles. Weight is updated as:
       \begin{equation}\label{12}
        \omega  = \omega _{\max }  - \left( {\frac{{\omega \max  - \omega _{\min } }}{{iter_{\max } }}} \right)iter
       \end{equation}
        where $\omega _{\min }$, $\omega _{\max }$, $iter$, and $iter_{\max }$   are minimum, maximum values of $\omega$ , the current iteration number and pre-specified maximum number of iteration cycles, respectively.

             \subsubsection{Proposed PSO-PID Controller}

This paper presents a PSO-PID controller for searching the
optimal controllers parameters of MIMO nonlinear system ,$kp_{i}$ , $ki_{i}$ and , $kd_{i}$
with the PSO algorithm. Each individual $K_{i}$ contains 3*m members
$kp_{i}$ , $ki_{i}$ and , $kd_{i}$. Its dimension is n*3*m.
The searching procedures of the proposed PSO-PID controller
were shown as below [12].
Optimal design for both conventional PID controllers can be fulfilled using PSO technique. Based on the
PSO technique, the PID controller can be tuned to some parameters values that minimize those fitness functions given in (15), (16) and(17).
The algorithmic steps for the PSO is as follows:
\begin{itemize}
  \item \textbf{Step} 1: Select the number of particles, generations, tuning accelerating coefficients $c_{1}$ and $c_{2}$ and random numbers $r_{1}$, and $r_{2}$ to start the optimal solution searching.
\end{itemize}

 \begin{itemize}
   \item \textbf{Step} 2: Initialize the particle position and velocity.
   \end{itemize}

  \begin{itemize}
    \item \textbf{Step} 3: Select the particle's individual best value for each
generation.
  \end{itemize}

\begin{itemize}
  \item \textbf{Step} 4: Select the particle's global best value, particle near the
target among all the particles, is obtained by comparing all the
individual best values.
\end{itemize}

\begin{itemize}
  \item \textbf{Step} 5: Update particle individual best pbest , global best gbest , in the velocity equation (18) and obtain the new velocity.
\end{itemize}

\begin{itemize}
  \item \textbf{Step} 6: Update the new velocity value in Eq. (19) and obtain the
position of the particle.
\end{itemize}

\begin{itemize}
  \item  \textbf{Step} 7: Find the optimal solution with a minimum (IAE, ISE, ITSE) from the
updated new velocity and position values.
\end{itemize}

\section{SIMULATION RESULTS}
This section presents a simulation  example to shown an application of the proposed  control algorithm and its satisfactory performance.
The MIMO nonlinear system is characterized by the equation(16), [16][17].
\begin{equation}\label{12}
   \left\{
    \begin{array}{ll}
        y1(k)=\frac{a_{1}y_{1}(k-1)y_{2}(k-1)}{1+a_{2}y_{1}^2(k-1)+a_{3}y_{2}^2(k-1)}+\\
        \,\,\,\,\,\,\,a_{4}u_{1}(k-2)+a_{5}u_{1}(k-1)+a_{6}u_{2}(k-1) \\
        y_{2}(k)=\frac{b_{1}y_{2}(k-1)sin(y_{2}(k-2))}{1+b_{2}y_{2}^2(k-1)+b_{3}y_{1}^2(k-1)}+\\
        \,\,\,\,\,\,\,b_{4}u_{2}(k-2)+b_{5}u_{2}(k-1)+b_{6}u_{1}(k-1)
    \end{array}
\right.
\end{equation}
The system parameters are: $a_{1}=0.7, a_{2}=1, a_{3}=1, a_{4}=0.3, a_{5}=1, a_{6}=0.2,
b_{1}=0.5, b_{2}=1, b_{3}=1, b_{4}=0.5, b_{5}=1 and b_{6}=0.2$
which is used as a test for control techniques introduced in this paper, to demonstrate the effectiveness of the proposed algorithms. Here $y_{1}$ and $y_{2}$  are the outputs, $u_{1}$ and $u_{2}$  are the inputs which is uniformly bounded in the region $[-2, 2]$.\\
We choose $[y_{1}(k-1),y_{1}(k-2),u_{1}(k-1),u_{1}(k-2),u_{2}(k-1)]$ and $[y_{2}(k-1),y_{2}(k-2),u_{2}(k-1),u_{2}(k-2),,u_{2}(k-1)]$ as inputs variables, and the number of fuzzy rules is four. The setup applied in this work was the following:  the population size was 20, the stopping criterion was 30 generations, $\omega_{min}=0.5$, $\omega_{max}=0.9$,  and $c_{1}= c_{2} =2$.\\

\begin{figure}[thpb]
      \centering
    \includegraphics[scale=0.47]{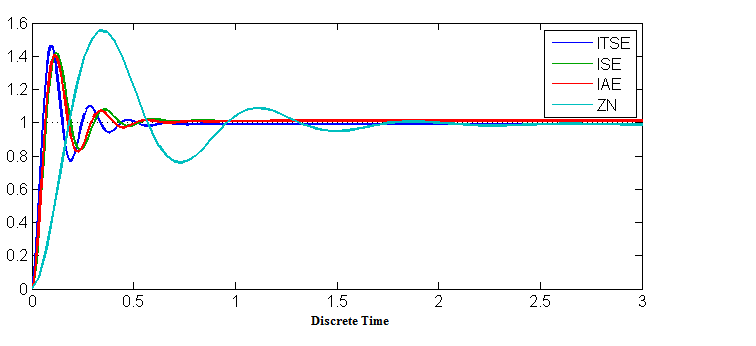}
       \caption{System response (y1)}
      \label{figure2}
   \end{figure}
\begin{figure}[thpb]
      \centering
    \includegraphics[scale=0.47]{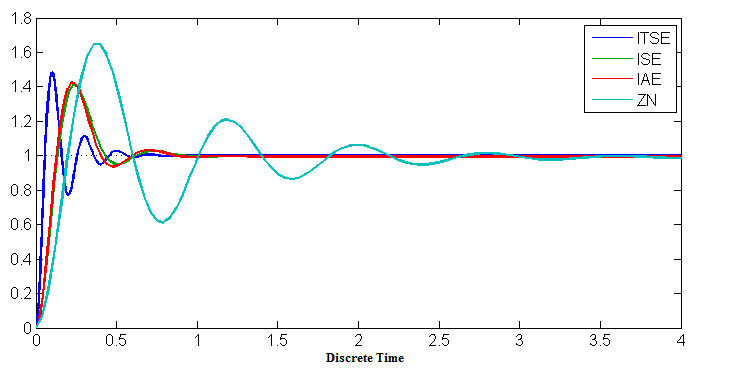}
       \caption{System response (y2)}
      \label{figure2}
   \end{figure}
   \begin{table}[h]
\caption{Optimized PID parameters (y1)}
\label{table_example}
\begin{center}
\begin{tabular}{|c||c||c||c|}
\hline
$\textbf{Tuning Method}$ & kp & ki & kd \\

\hline
\textbf{Z-N-PID} & 32.3431 & 3.2943&4.4829\\
\hline
\textbf{PSO-PID1} (ISE) &$26.7236$ & 46.2192&30.136 \\
\hline
\textbf{PSO-PID1 (IAE)} &$8.7263$ & 46.2192&33.1360 \\
\hline
\textbf{PSO-PID1 (ITSE)}&$5.3487$ & 15.7988&45.4369 \\
\hline
\end{tabular}
\end{center}
\end{table}
\begin{table}[h]
\caption{Optimized PID parameters (y2)}
\label{table_example}
\begin{center}
\begin{tabular}{|c||c||c||c|}
\hline
$\textbf{Tuning Method}$ & kp & ki & kd \\

\hline
\textbf{Z-N PID} & 38.0743 & 4.6739&5.6734\\
\hline
\textbf{PSO-PID1 (ISE)} &$35.7236$ &16.2192&9.136\\
\hline
\textbf{PSO-PID2 (IAE)} &$39.5246$ &17.2514&9.7983 \\
\hline
\textbf{PSO-PID3 (ITSE)}&$45.3487$ &15.7988&41.4369 \\
\hline
\end{tabular}
\end{center}
\end{table}
\begin{table}[h]
\caption{Step response performance for PID controllers (y1)}
\label{table_example}
\begin{center}
\begin{tabular}{|c||c||c||c|}
\hline
$\textbf{Tuning Method}$ & \textbf{Overshoot}(\%) & \textbf{Rise Time} & \textbf{Setting Time}\\
\hline
\textbf{Z-N PID} & 55.3483& 0.1264&1.6733\\
\hline
\textbf{SPSO-PID1 (ISE)} &$41.6977$ & 0.0474& 0.6182 \\
\hline
\textbf{PSO-PID2 (IAE)} &$40.5825$ & 0.0453&0.4791 \\
\hline
\textbf{PSO-PID3 (ITSE)}t&$46.1849$ & 0.0374&0.4246 \\
\hline
\end{tabular}
\end{center}
\end{table}
\begin{table}[h]
\caption{Step response performance for PID controllers (y2)}
\label{table_example}
\begin{center}
\begin{tabular}{|c||c||c||c|}
\hline
$\textbf{Tuning Method}$ & \textbf{Overshoot(\%)} & \textbf{Rise Time} & \textbf{Setting Time}\\
\hline
\textbf{Z-N PID} & 64.8174& 0.1352&3.2941\\
\hline
\textbf{PSO-PID1 (ISE)} &$40.8062$ &  0.0929& 0.7970 \\
\hline
\textbf{PSO-PID2 (IAE)} &$42.3434$ &0.0884&0.7775 \\
\hline
\textbf{PSO-PID3 (ITSE)}&$48.5050$ & 0.0389&0.5265\\
\hline
\end{tabular}
\end{center}
\end{table}
\newpage
In the conventionally Z-N tuned PID controller,  the systems
response produces high overshoot, but a better performance
obtained with the implementation of PSO-based PID
controller tuning. In the PSO-based PID controllers (PSO-PID),
different performance index gives different results.
These are shown in TABLE. III and TABLE. IV. Comparative results for the PID controllers are given
below in TABLE. V and TABLE. VI where the step response performance is evaluated based on the overshoot, settling time and Rise time.
The corresponding plot for the step responses are shown in Fig. 2 and Fig.3.

%
\section{CONCLUSIONS}
This paper presents a  design method for determining the PID controller parameters using the PSO method for MIMO nonlinear systems. The proposed method integrates the PSO algorithm with  performance criterions into a PSO-PID controller. The comparison between PSO-based PID (PSO-PID) performance and the ZN-PID is presented. The results show the advantage of
the PID tuning using PSO-based optimization approach.


\end{document}